\documentclass[aps,amssymb,amsmath,prl,reprint,noshowpacs,]{revtex4-1}
\usepackage{times}
\usepackage{amssymb,amsmath,graphicx}
\usepackage[usenames]{color}
\usepackage{bbold}

\newcommand{\imu}{\text{\rm i}}
\usepackage{textcomp}
\newcommand{\murm}{\hbox{\textmu}}

\newcommand{\figwidth}{0.95\columnwidth} 
\newcommand{\commentOut}[1]{}

\begin{document}
\title{Spontaneous Emission Control in a Tunable Hybrid Photonic System}
\author{Martin Frimmer}
\affiliation{Center for Nanophotonics, FOM Institute AMOLF,
Science Park 104, 1098 XG Amsterdam, The Netherlands}
\email{frimmer@amolf.nl}
\author{A. Femius Koenderink}
\affiliation{Center for Nanophotonics, FOM Institute AMOLF,
Science Park 104, 1098 XG Amsterdam, The Netherlands}

\begin{abstract}
We experimentally demonstrate control of the rate of spontaneous emission in a tunable hybrid photonic system that consists of two canonical building blocks for spontaneous emission control, an optical antenna and a mirror, each providing a modification of the local density of optical states (LDOS).
We couple fluorophores to a plasmonic antenna to create a superemitter with an enhanced decay rate. In a superemitter analog of the seminal Drexhage experiment we probe the LDOS of a nanomechanically approached mirror. Due to the electrodynamic interaction of the antenna with its own mirror image the superemitter traces the inverse LDOS of the mirror, in stark contrast to a bare source, whose decay rate is proportional to the mirror LDOS\@.
\end{abstract}
\date\today

\maketitle

\begin{figure}
\includegraphics[width=\figwidth]{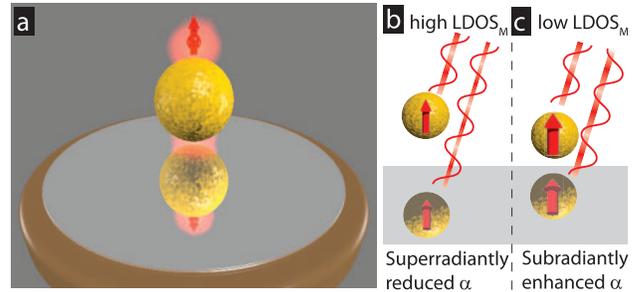}
\caption{(a)~A fluorescent source and an optical antenna form a superemitter. In close proximity of a mirror the antenna particle electrodynamically interacts with its mirror image.
(b)~When located at a position of enhanced mirror $\text{LDOS}_\text{M}$ the field scattered by the antenna interferes constructively with radiation emerging from its own mirror image. This superradiant damping reflects in a reduced antenna polarizability $\alpha$.
(c)~At a position of reduced $\text{LDOS}_\text{M}$ destructive interference leads to subradiantly enhanced $\alpha$.}
\label{chSEexperiment:fig:Cartoon}
\end{figure}

The study of light-matter interaction is a cornerstone of contemporary physics.
According to Fermi's golden rule the rate of spontaneous emission of light can be controlled through engineering of photonic modes~\cite{Novotny2006}.
In a seminal experiment Drexhage modified the decay rate of fluorophores by varying their distance to a mirror~\cite{Drexhage1970}.
Already in 1946 Purcell had suggested to boost the decay rate of a source by coupling it to a resonant cavity~\cite{Purcell1946}.
Both Drexhage's and Purcell's works are nowadays discussed in terms of the local density of optical states (LDOS) a quantity governing spontaneous emission, thermal radiation, and vacuum mediated forces~\cite{Novotny2006}.
A rich toolbox of photonic systems to control various aspects of spontaneous emission has been established, including cavities~\cite{Vahala2005}, mirrors~\cite{Snoeks1995,Buchler2005} and photonic crystals~\cite{Noda2007} that all shape the LDOS on a wavelength scale. Length scales even smaller than the wavelength are the realm of nanophotonics~\cite{Novotny2006} whose prototypical building block for spontaneous-emission control is the optical antenna, which exploits plasmonic resonances of metal nanoparticles~\cite{Novotny2011}. By coupling a source of spontaneous emission to such a nano-antenna, a `superemitter'~\cite{Farahani2005} retaining the dipolar nature of the source yet exhibiting a boosted decay rate can be created~\cite{Kuehn2006}.
Currently, nanophotonics is combining and integrating these functional units into `hybrid photonic systems' in order to boost figures of merit by embedding nanoplasmonic elements in cavities, stratified media or photonic crystals~\cite{Benson2011,Chen2012}.
Importantly, in such a photonic hybrid the building blocks are expected to interact with each other, such that the $\text{LDOS}_\text{hyb}$ of the hybrid emerges from the respective LDOS of the individual building blocks in a non-trivial fashion~\cite{Frimmer2012a}.
Consider a superemitter in front of a mirror, as sketched in Fig.~\ref{chSEexperiment:fig:Cartoon}(a), where the mirror represents a background system in which the superemitter is immersed. In this superemitter-equivalent of the classic Drexhage experiment a nano-antenna provides a rate enhancement $\text{LDOS}_\text{ant}$ to a fluorophore, e.g. a molecule. While the approaching mirror itself certainly provides an enhancement $\text{LDOS}_\text{M}$ to the source, the antenna particle will interact with its mirror image, such that the mirror clearly not only acts on the source itself but furthermore can effectively modify the antenna. Therefore, a tunable photonic background system should offer a route to dynamically mould the properties of an optical antenna and thereby the rate enhancement it provides to a spontaneous emitter.
However, the emergence of such a hybrid photonic LDOS from the individual LDOS of the constituents has never been unraveled to date.

This Letter experimentally demonstrates the dynamic control of the rate enhancement provided by an optical antenna  to a spontaneous emitter in a hybrid photonic system. By nanomechanically approaching a metallic mirror to the source-antenna ensemble we perform the superemitter equivalent of the classic Drexhage experiment.
We find that the superemitter probes the \emph{inverse} $\text{LDOS}_\text{M}$ of the mirror which is caused by the strong interaction between the optical antenna and its own mirror image.
\begin{figure}
\includegraphics[width=\figwidth]{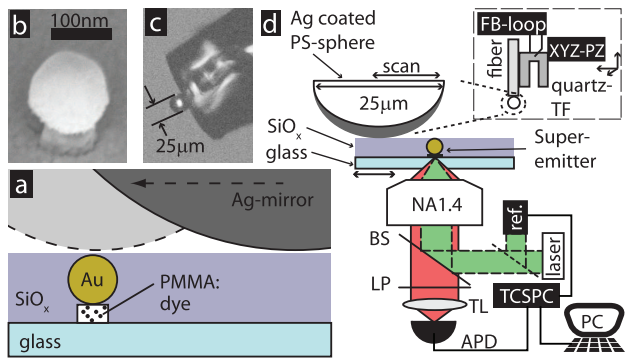}
\caption{
(a)~Experimental principle: The distance between a superemitter and a spherical mirror is varied by scanning the mirror-sphere laterally across the sample.
(b)~SEM micrograph of fabricated superemitter, composed of a Au nanoparticle residing on a dye-doped PMMA pedestal on a glass substrate. For SEM imaging a 5\,nm Au layer was sputtered.
(c)~Photograph of micromirror on cleaved optical fiber.
(d)~Experimental setup: The sample is located on an inverted microscope with the micromirror on the fiber located above. The fiber is attached to a quartz tuning fork, which can be positioned with an xyz-piezo [see inset]. A beamsplitter (BS) focuses the pump laser via the objective on the sample. Fluorescence filtered by a long-pass filter (LP) is imaged by the tube-lens (TL) on the detector, which can be an APD, a CCD, or a spectrometer.
}
\label{chSEexperiment:fig:setup}
\end{figure}

\begin{figure*}
\includegraphics[width=\textwidth]{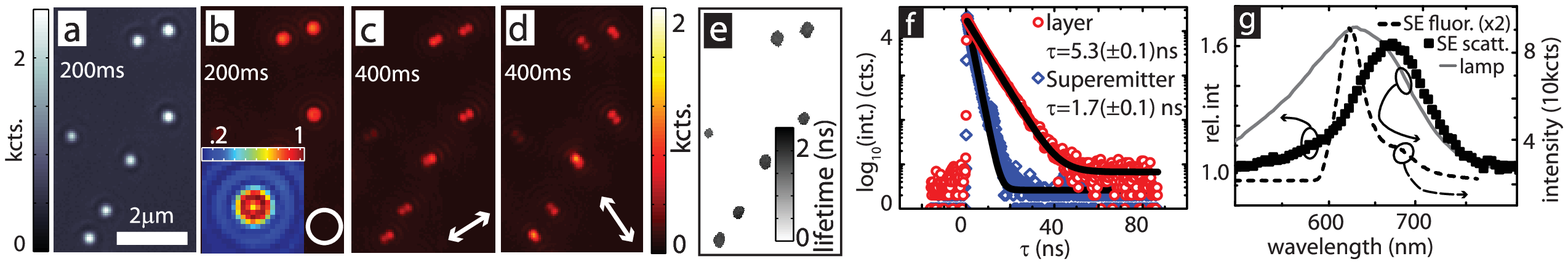}
\caption{
(a)~CCD image of superemitters under white-light illumination, showing antenna particles as bright scatterers.
(b)~Unpolarized fluorescence-intensity CCD image of area in~(a) under circularly polarized epi-illumination. Each superemitter appears as a bright source of fluorescence. Acquisition time 200\,ms. Inset: Enlarged view of normalized intensity of superemitter showing donut shape.
(c)~Same as~(a) but with linear polarizer in detection path along direction indicated by white arrow. Acquisition time 400\,ms.
(d)~Same as~(c) but with polarizer rotated by 90$^\circ$.
(e)~Fluorescence lifetime image of sample area in~(a--d), showing lifetime of superemitters between 1.5 and 2\,ns.
(f)~Decay traces of bare dye layer (circles) and fabricated superemitter (diamonds), both fit with a single-exponential decay (solid lines), yielding a lifetime of 5.3\,ns for the layer and 1.7\,ns for the superemitter.
(g)~Black squares: Scattering spectrum of an antenna particle normalized by emission spectrum of white-light source (solid line). Dashed line: Superemitter fluorescence spectrum.}
\label{chSEexperiment:fig:Superemitter_Char}
\end{figure*}

For our experiments we assembled superemitters [see sketch in Fig.~\ref{chSEexperiment:fig:setup}(a)] by co-localizing fluorescing dye molecules with near-unity quantum yield (Bodipy TR, Invitrogen) with strongly scattering Au colloids (diameter 100\,nm, BBInternational). A dye-doped PMMA layer (60\,nm thickness) is spin coated on a glass cover slip~\cite{Frimmer2012b} onto which Au-colloids in solution are spin coated. An oxygen plasma removes PMMA and embedded fluorophores from the sample surface, except where the Au particle acts as an etch mask~\cite{Sorger2011}. As a result, we obtain isolated Au particles residing on dye-doped PMMA pedestals [diameter ca.\ 70\,nm, see Fig.~\ref{chSEexperiment:fig:setup}(b)]. We estimate the pedestal to contain several hundred dye molecules. The size of the pedestal renders the effect of quenching negligible, which only plays a role at emitter-metal distances $<$10\,nm~\cite{Anger2006}. Finally, we cover the sample with a layer of about 120\,nm spin-on glass (FOX-14, Dow Corning) for mechanical protection.

We characterize the superemitters optically in the setup reported in Ref.~\onlinecite{Frimmer2011} and sketched in Fig.~\ref{chSEexperiment:fig:setup}(d).
Under white light illumination the Au particles appear as bright scatterers on a CCD camera, as shown in Fig.~\ref{chSEexperiment:fig:Superemitter_Char}(a).
Under epi-illumination by a circularly polarized pump laser (532\,nm, repetition rate 10\,MHz, pulse width $<$10\,ps) the fluorescence image of the region in Fig.~\ref{chSEexperiment:fig:Superemitter_Char}(a) appears as shown in Fig.~\ref{chSEexperiment:fig:Superemitter_Char}(b).
Clearly, fluorescence emerges where scatterers are located.
Since the molecules are immobilized under the Au particle we expect that the dipole moment induced in the optical antenna, which dominates emission of the superemitter, is oriented along the optical axis~\cite{Taminiau2008b}. Accordingly, in fluorescence, superemitters appear as donut-shaped patterns on the CCD [see inset of Fig.~\ref{chSEexperiment:fig:Superemitter_Char}(b)]~\cite{Lieb2004}.
Furthermore, filtering the signal that led to Fig.~\ref{chSEexperiment:fig:Superemitter_Char}(b) with a linear polarizer in the detection path yields Fig.~\ref{chSEexperiment:fig:Superemitter_Char}(c), which exhibits the expected double-lobed pattern~\cite{Lieb2004}, which is furthermore practically unchanged in intensity and follows the polarizer axis when the analyzer is rotated [see Fig.~\ref{chSEexperiment:fig:Superemitter_Char}(d)].
To characterize the decay rate of our superemitters Fig.~\ref{chSEexperiment:fig:Superemitter_Char}(e) shows a fluorescence lifetime image of the area investigated for Figs.~\ref{chSEexperiment:fig:Superemitter_Char}(a-d), where we have clamped the lifetime value of pixels holding less than 1000\,events to zero. The distribution of lifetimes exhibited by the superemitters ranges from about 1.5 to 2\,ns. An example for the decay behavior of a typical superemitter is shown in Fig.~\ref{chSEexperiment:fig:Superemitter_Char}(f). The decay of the superemitter (blue diamonds) is fitted well with a single exponential with time constant 1.7\,ns.
To judge the enhancement provided by the Au particle the superemitter lifetime has to be compared to the lifetime of the dye molecules in absence of the antenna. To this end, we measure the decay in a reference section of the sample where no Au particles are present and which has been protected from the plasma etch. The bare dye molecules decay single exponentially with time constant 5.3\,ns, shown as the open circles in Fig.~\ref{chSEexperiment:fig:Superemitter_Char}(f).

The observed rate enhancement is a result of the plasmonic resonance of the Au nanoparticle~\cite{Kuehn2006,Mertens2007}. To characterize the spectral matching of emitter and antenna we analyze the emission spectrum of the superemitters, shown as the dashed line in Fig.~\ref{chSEexperiment:fig:Superemitter_Char}(g). The superemitter emission peaks around 620\,nm and is broadened by a shoulder to span up to about 700\,nm in close resemblance to the spectrum of the incorporated dye~\cite{Johnson2010}. To characterize the antenna particle we show a typical superemitter scattering spectrum as the black squares in Fig.~\ref{chSEexperiment:fig:Superemitter_Char}(g), where the spectrum of the used white-light source [grey line] has been normalized out. The particle's scattering spectrum [black squares in Fig.~\ref{chSEexperiment:fig:Superemitter_Char}(g)] exhibits a resonant line-shape, peaking around 665\,nm and spanning a width of about 70\,nm, while well overlapping the superemitter emission spectrum.

We now turn to our nanomechanical version of Drexhage's experiment~\cite{Drexhage1970} whose implementation is inspired by the method developed by Buchler \emph{et al.}~\cite{Buchler2005} and relies on moving a spherical micromirror attached to a scanning probe~\cite{Frimmer2012b}. As long as the diameter of the spherical mirror largely exceeds its distance to the source it serves as a good approximation of a flat mirror. Our scheme of changing the distance between a fluorophore and the mirror is illustrated in Fig.~\ref{chSEexperiment:fig:setup}(a). The fluorescing source is fixed in a substrate while a large spherical mirror with a diameter of $25\,\murm$m is laterally moved across the sample surface.
The mirror-sample distance is kept constant at ca.\ 5\,nm with a shear-force feedback loop~\cite{Novotny2006}.
In Fig.~\ref{chSEexperiment:fig:setup}(a) two positions of the mirror with respect to a superemitter are shown to illustrate the principle of changing the emitter-mirror distance. The micromirror (a polystyrene bead covered with a 400\,nm evaporated Ag layer) is glued to a cleaved optical fiber as shown in Fig.~\ref{chSEexperiment:fig:setup}(c). The optical fiber is then glued to a quartz tuning fork, as sketched in Fig.~\ref{chSEexperiment:fig:setup}(d, inset).
Measurements on single emitters have confirmed that our method indeed exactly replicates Drexhage's calibrated LDOS experiment~\cite{Frimmer2012b}.

\begin{figure}
\includegraphics[width=\figwidth]{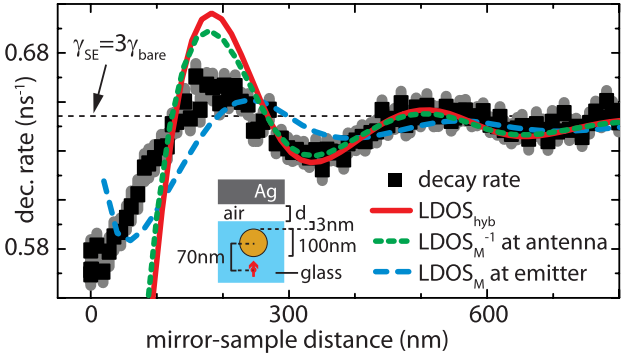}
\caption{Black squares: Decay rate of superemitter as a function of mirror-sample distance (error-bars denote $1\sigma$). Horizontal dashed line: Superemitter decay rate $\gamma_\text{SE}$, i.e., 3-fold antenna-enhanced rate of bare emitters $\gamma_\text{bare}$. Red line: Calculated enhancement $\text{LDOS}_\text{hyb}$ for vertically oriented source coupled to antenna (see inset for geometry) as a function of air-gap thickness. Calculated enhancement follows $\text{LDOS}_\text{M}^{-1}$ at antenna position (green dotted line). Blue dashed line: Calculated $\text{LDOS}_\text{M}$ at source position in absence of antenna.}
\label{chSEexperiment:fig:Superemitter_Drex}
\end{figure}

To perform the superemitter-equivalent of Drexhage's experiment we combine our two methods to control spontaneous emission and approach the mirror to a superemitter, thereby dynamically tuning the LDOS experienced by the source-antenna assembly. To this end, we raster-scan our micromirror over a superemitter while continuously measuring its lifetime. As a result, we obtain the decay rate of the superemitter as a function of mirror-sample distance, shown as the black squares in Fig.~\ref{chSEexperiment:fig:Superemitter_Drex}. There is a clear variation visible in the decay rate as a function of mirror-sample separation. Note that this variation is due to the LDOS modification provided by the mirror on top of the three-fold enhancement provided by the optical antenna.

To model our experiments, we consider an air layer sandwiched between two semi-infinite half-spaces [see sketch in inset of Fig.~\ref{chSEexperiment:fig:Superemitter_Drex}], the upper one being Ag ($\epsilon=-15.5+0.52\imu$ at 620\,nm, measured by ellipsometry on an Ag film on a Si substrate), the lower one being glass ($\epsilon=2.25$) and unity quantum yield of the emitters~\cite{Frimmer2012b}.
We furthermore take into account that our superemitters are strongly polarized along the optical axis as established from Figs.~\ref{chSEexperiment:fig:Superemitter_Char}(b-d). We therefore consider an emitter oriented perpendicularly to the interface and located at a fixed depth of 123\,nm in the glass substrate as sketched in the inset of Fig.~\ref{chSEexperiment:fig:Superemitter_Drex}. The optical antenna is described in a dipole model as a polarizable sphere of 100\,nm diameter whose center is located 53\,nm into the glass substrate~\footnote{
The polarizability of the antenna particle is described in a Drude model with characteristic frequency $\omega_0=3.4\times10^{15}\,\text{s}^{-1}$ and damping rate $\gamma=8.5\times10^{12}\,\text{s}^{-1}$ to match it to the scattering spectrum in Fig.~\ref{chSEexperiment:fig:Superemitter_Char}(g).}.
Our model is fully analytical and electrodynamic, taking into account the full multiple-scattering process between the antenna and the double interface~\cite{Frimmer2012a}.

The red solid line in Fig.~\ref{chSEexperiment:fig:Superemitter_Drex} is the calculated decay-rate enhancement $\text{LDOS}_\text{hyb}$ experienced by the source coupled to the polarizable particle in front of the mirror (scaled with the antenna-enhanced decay rate of the source) as a function of air-gap thickness. The calculated hybrid enhancement $\text{LDOS}_\text{hyb}$ is in excellent quantitative agreement with the measured data for mirror-sample distances larger than ca.\ 280\,nm. At distances smaller than 280\,nm\ there is good qualitative agreement between measurement and calculation while the measured decay-rate modifications are smaller than those theoretically predicted.
We also plot the \emph{inverse} $\text{LDOS}_\text{M}$ of the mirror at the position of the antenna particle (scaled with the antenna-enhanced decay rate of the source) as a function of mirror-sample separation as the green dotted line in Fig.~\ref{chSEexperiment:fig:Superemitter_Drex}. Clearly, the enhancement calculated for the superemitter [red solid line in Fig.~\ref{chSEexperiment:fig:Superemitter_Drex}] closely follows the inverse $\text{LDOS}_\text{M}$ in front of the mirror.
The \emph{inverse} proportionality of the superemitter enhancement factor to the mirror $\text{LDOS}_\text{M}$ reflects that the antenna, which dominates the decay of the superemitter, behaves very differently from a quantum emitter, whose decay rate is \emph{proportional} to the LDOS\@.
This stark contrast is a result of the antenna being a strong scatterer driven by the source and subjected to its own scattered field~\cite{Frimmer2012a} as opposed to a quantum emitter corresponding to a constant current source in a classical treatment~\cite{Novotny2006}.
At positions of enhanced mirror $\text{LDOS}_\text{M}$ the scattered field, reflected from the mirror and arriving back at the antenna with a phase difference, effectively depolarizes the antenna. More intuitively stated, as illustrated in Figs.~\ref{chSEexperiment:fig:Cartoon}(b,c), when the mirror $\text{LDOS}_\text{M}$ is high (low) the scattered field of the antenna interferes constructively (destructively) with that of its own mirror image, effectively forming a super- (sub-)radiant hybrid plasmonic mode~\cite{Prodan2003,Buchler2005}.
For a subradiant mode, the suppressed radiative damping results in an enhanced antenna resonance strength.
Accordingly, the decay rate of a superemitter can be boosted beyond $\text{LDOS}_\text{ant}$ by embedding it in a background system with reduced LDOS, ideally approaching a photonic bandgap~\cite{Noda2007}.
This counterintuitive inverse LDOS effect is generic for any antenna whose damping rate is mainly radiative~\cite{Frimmer2012a}, i.e., for any large antenna with scattering cross section close to the upper bound $\sigma_s={3/2\pi}\lambda^2$ known as the unitary limit.

Finally, to exclude that the measured variation of decay rate is the result of the dye molecules themselves experiencing the $\text{LDOS}_\text{M}$ of the approaching mirror we also plot $\text{LDOS}_\text{M}$ (scaled with the antenna-enhanced decay rate of the source) at the position of the source as the blue dashed line in Fig.~\ref{chSEexperiment:fig:Superemitter_Drex}. Clearly, the $\text{LDOS}_\text{M}$ at the source itself is incommensurable with the measured data.
Figure~\ref{chSEexperiment:fig:Superemitter_Drex} therefore represents the experimental confirmation that an increased background LDOS indeed \emph{reduces} the enhancement experienced by a source coupled to a strongly scattering antenna.
Regarding the discrepancy between the calculation and the measurement for very small mirror-sample distances in Fig.~\ref{chSEexperiment:fig:Superemitter_Drex} we speculate that the finite size of our antenna particle starts to play a role on such small length scales. This regime offers the exciting prospect of engineering higher-order multipolar analogues of the LDOS~\cite{Chen2012,Andersen2011}.

In conclusion, we have coupled spontaneous emitters confined to a subwavelength volume to an optical antenna, creating a superemitter exhibiting a decay-rate enhancement of three.
We actively tuned the rate enhancement provided by the antenna by nanomechanically approaching a mirror to the superemitter. Importantly, we found that the decay-rate enhancement $\text{LDOS}_\text{hyb}$ experienced by the source in the superemitter varies in proportion to the inverse LDOS of the mirror~\cite{Frimmer2012a} as a result of the antenna effectively hybridizing with its own mirror image.
Our system is inherently broadband, since the resonance of the optical antenna is broad and the mirror also has no characteristic resonance. It will be most interesting to extend our study to emitters coupled to both deep-subwavelength optical antennas and super-wavelength resonators, like microspheres or microtoroids~\cite{Vahala2005}, which can be manipulated nanomechanically as well~\cite{Mazzei2007}.
Furthermore, our results shed new light on approaches to efficiently interface single emitters via optical antennas with microresonators~\cite{Xiao2012} or waveguides~\cite{BernalArango2012}.

\begin{acknowledgments}
This work is part of the research program of the ``Stichting voor Fundamenteel
Onderzoek der Materie (FOM)'', which is financially supported by the
``Nederlandse Organisatie voor Wetenschappelijk Onderzoek (NWO)''.
AFK gratefully acknowledges an NWO-Vidi grant for financial support.
\end{acknowledgments}

\bibliography{Frimmer_bib}

\end{document}